\documentclass[a4paper,12pt,notoc]{JHEP3}

\usepackage{latexsym}
\usepackage{amsmath,amssymb,amsxtra,amscd,amsthm}
\usepackage[mathscr]{eucal}

\DeclareFontFamily{U}{rsfs}{\skewchar\font127 }
\DeclareFontShape{U}{rsfs}{m}{n}{
   <5> rsfs5
   <6> rsfs6
   <7> rsfs7
   <8> rsfs8
   <9> rsfs9
   <10> rsfs10
   <10.95> rsfs11
   <12> rsfs12
   <14.4> rsfs14
   <17.28> rsfs17
   <20.74> rsfs20
   <24.88> rsfs25
   <29.86-> rsfs30}{}
\DeclareMathAlphabet\scr{U}{rsfs}{m}{n}

\def\cN{\mathcal{N}}

\def\sF{\mathscr{F}}

\def\sO{\mathscr{O}}

\def\mC{\mathbb{C}}

\def\mF{\mathbb{F}}

\def\mN{\mathbb{N}}

\def\mP{\mathbb{P}}

\def\mZ{\mathbb{Z}}

\DeclareMathOperator{\chern}{ch}
\DeclareMathOperator{\ch}{c}
\DeclareMathOperator{\todd}{td}

\DeclareMathOperator{\rank}{rk}

\DeclareMathOperator{\Ext}{Ext}

\DeclareMathOperator{\End}{End}
\def\bb1{\textup{\small{1}} \kern-3.8pt \textup{1}}

\newcommand{\e}[1]{e^{#1}}

\newcommand{\lcm}[1]{\textrm{lcm} (#1)}
\newcommand{\BS}[1]{| #1 \rangle\!\rangle_B}

\title{On D0-branes in Gepner Models}
\author{Emanuel Scheidegger\thanks{Research supported by the German-Israeli Foundation for Scientific Research and Development}\\
\textit{Max-Planck Institut f\"ur Gravitationsphysik}\\
\textit{Albert-Einstein-Institut}\\
\textit{Am M\"uhlenberg 1}\\
\textit{14776 Golm, Germany}\\
\email{esche@aei.mpg.de}}
\abstract{We show why and when D0-branes at the Gepner point of Calabi-Yau manifolds given as Fermat hypersurfaces exist. 

}
\preprint{hep-th/0109013\\AEI-2001-109}
\keywords{Superstrings and Heterotic Strings, D-branes, Conformal Field Models in String Theory, Differential and Algebraic Geometry}

\begin{document}

\section{Introduction}
\label{sec:Introduction}

D-branes in Calabi-Yau compactifications are still not as well understood as D-branes in flat spaces or toroidal compactifications, although there have recently been many results shedding more and more light on this aspect of string theory. Thereby new and interesting relationships between physics and mathematics have been discovered. For recent reviews see~\cite{Douglas:2000be}, \cite{Douglas:2001hw} and~\cite{Govindarajan:2001jc}. 

One interesting and puzzling aspect has been the existence of the D0-brane at the Gepner point in the K\"ahler moduli space of a Calabi-Yau compactification in Type II string theory. We briefly review some facts about the D0-brane. It is seemingly the simplest D-brane and we expect it to be in the spectrum and to have a moduli space of dimension 3, since it can move everywhere on the Calabi-Yau manifold which is therefore its moduli space. In terms of the categorical formulation of the classification of D-branes on Calabi-Yau manifolds given in~\cite{Douglas:2000gi}, \cite{Diaconescu:2001ze} it is an object of the derived category of coherent sheaves on $X$ which contains not only the information about its charge but also about its location on $X$. This has been made more precise in~\cite{Aspinwall:2001pu}. According to the postulated equivalence of this derived category to the derived category of representations of certain quivers which describes the D-brane spectrum at small volume, there should also be an object corresponding to the D0-brane in the Gepner model of $X$. D-branes in Gepner models are described by rational boundary states~\cite{Recknagel:1998sb}. 

In~\cite{Brunner:2000jq} it has been observed that the D0-brane was not in the D-brane spectrum of the rational boundary states in the quintic. This absence can be argued to be consistent with the geometric hypothesis as follows~\cite{Douglas:2000vm}. Any location we might pick for the D0-brane would break some of the symmetry group $\mZ_5^4$, but all of the rational B-type boundary states are singlets under this group and hence we should not find the D0-brane in this analysis. However, in~\cite{Diaconescu:2000vp} and~\cite{Scheidegger:2000ed} the D0-brane was found in the spectrum of other Gepner models. We will argue that this observation is not in contradiction to the symmetry argument. In contrast, this argument can be used to predict the existence of D0-branes at the Gepner point.

Therefore we will review in Section~\ref{sec:Symmetries} the symmetries of Gepner models and the corresponding Fermat hypersurfaces in weighted projective spaces. In Section~\ref{sec:Boundary_states} we will first summarize the necessary results on boundary states in Gepner models. Subsequently, we will argue the existence of D0-branes as rational boundary states. In the special case where the hypersurface is an elliptic fibration we can say more about their existence due to the additional geometric structure. This will be the subject of Section~\ref{sec:Elliptic_fibrations}. Some consequences on walls of marginal stability are discussed in Section~\ref{sec:Lines}. We end in Section~\ref{sec:Conclusions} with some conclusions.

\section{Symmetries of the Gepner Model}
\label{sec:Symmetries}

The Gepner model is a tensor product of $r$ $\cN=2$ minimal models at levels $k_j$, $j=1,\dots,r$ whose central charges $c_j=\frac{3k_j}{k_j+2}$ add up to $9$ (for a string theory compactification down to four dimensions), subject to the GSO projection to guarantee space-time supersymmetry and augmented by twisted sectors to restore modular invariance. We will discuss in this section the symmetries of Gepner models with $r=4$ and $r=5$ A-type modular invariants and the corresponding Fermat hypersurfaces in weighted projective spaces. Recall that to each minimal model factor at level $k_j$ with A-type modular invariant there is an associated monomial in the defining polynomial $W$ of the Calabi-Yau hypersurface $X$ of the form $z_j^{k_j+2}$, where $z_j$ is a homogeneous coordinate of the weighted projective space $\mP^4_w$ with weights $w=(w_1,\dots,w_5)$. Since we need for a three-dimensional hypersurface in a four-dimensional weighted projective space five homogeneous coordinates, we add a free minimal model factor theory with A-type modular invariant at level $k_5=0$ in case that the Gepner model consists only of four factors.

Depending on whether there are four or five factor theories in the Gepner model the relation between the symmetry groups of the Gepner models and those of the Calabi-Yau manifold $X$ is very different. It is most convenient to describe the symmetries of the Gepner models using the language of simple currents. (For a general review, see~\cite{Schellekens:1990xy}; for applications to boundary states in Gepner models see~\cite{Brunner:2000nk}, \cite{Fuchs:2000gv}, \cite{Brunner:2000wx} and~\cite{Fuchs:2001fd}). We will follow the notation of~\cite{Fuchs:2000gv}. Among the simple currents in the $j^\mathrm{th}$ $\cN=2$ minimal model we have $p_j$, $v_j$ and $s_j$. Denote the total spinor current by $s=(s_1,\dots,s_r)$ and define $v=(v_1,1,\dots,1)$. The internal part of the Gepner model (the Calabi-Yau extension in~\cite{Fuchs:2000gv}) is then obtained as a simple current extension of the tensor product of the minimal models by the group $G_{\textrm{CY}} = \langle \mathrm{u}, \mathrm{w}_j \rangle$. Here, $\mathrm{w}_j = v_1v_j,\;j=2,\dots r$ impose the fermion alignment and $\mathrm{u}=s^2v$ the internal part of the GSO projection. The order of $\mathrm{u}$ is $K=\lcm{k_j+2}$. Modulo identifications by the alignment currents $\mathrm{w}_i$, we may rewrite $\mathrm{u}$ in terms of the phase symmetries as $\mathrm{u} = \prod_{i=1}^r p_i v^{n+r}$, where $n$ is the dimension of $X$.

Since D0-branes satisfy B-type boundary conditions, they are described by B-type boundary states which will be the focus of Section~\ref{sec:Boundary_states}. These are constructed as A-type boundary states on the mirror. In the following we will study some properties of the mirror which will be needed later on. The mirror is obtained from the tensor product via the Greene-Plesser construction which can be reformulated as a simple current extension~\cite{Fuchs:2001fd} by the group 
\begin{equation}
  \label{eq:GPgroup}
  G_{\textrm{mirr}} = \left\langle \mathrm{w}_j, j=2,\dots,r; v_1^\varepsilon\prod_{j=1}^r p_j^{\alpha_j}, \frac{\varepsilon}{2}-\sum_{j=1}^r\frac{\alpha_j}{k_j+2}=0 \mod \mZ \right\rangle
\end{equation}
with $\varepsilon = 0$ for $r=5$ and $\varepsilon = 0,1$ for $r=4$. The Greene-Plesser group $G_{\textrm{GP}}$ is the quotient of $G_{\textrm{mirr}}$ by the group $\mZ_N$. $N$ depends on whether $\mathrm{u}$ is contained in~\eqref{eq:GPgroup} or not. The latter happens for $r=4$. Since $\mathrm{u}^2$ is always contained in~\eqref{eq:GPgroup}, $N=K$ for $r=5$ and $N=\frac{K}{2}$ for $r=4$. 

In the case $r=4$, we may formally treat the simple current $v_1$ as a phase symmetry $p_5$ and write $\alpha_5$ instead of $\varepsilon$. This corresponds to formally adding a trivial minimal model factor with $k_5=0$. Since such a theory is trivial, it does not have any symmetries at all, although formally one could associate to it a $\mZ_2$ phase symmetry, $p_5$. This is actually taken over by the simple current $v_1$. Viewing $v_1$ as a phase symmetry, $G_{\textrm{mirr}}$ takes the same form for both, $r=4$ and $r=5$. Only for $r=4$ however, $G_{\textrm{GP}}$ will, as an abstract abelian group, always contain a $\mZ_2$ factor. At the end of this section we will give a geometric picture of this reinterpretation of a vector symmetry as a phase symmetry. We will return to it in the next section when we discuss the labeling of the B-type boundary states. As we will see, this $\mZ_2$ factor is the origin for the existence of the D0-brane.

Viewed as an abstract abelian group, $G_{\textrm{mirr}}$ is a quotient by a $\mZ_K$. This quotient entails a $\mZ_K$ quantum symmetry and induces an enhanced discrete symmetry at the Gepner point of the K\"ahler moduli space of the corresponding Calabi-Yau manifold $X$~\cite{Vafa:1989ih}. This symmetry acts on the fields $\Phi_j$ in the NS sector of the Gepner model or the LG orbifold theory as
\begin{equation}
  \label{eq:quantum_symmetry}
  g(\Phi_j) = \e{\frac{2\pi i}{k_j+2}}\Phi_j = \e{2\pi i \frac{w_j}{d}}\Phi_j \qquad j=1,\dots,r
\end{equation}
where we have used that $w_j=\frac{d}{k_j+2}$ and $d=K$. The second equation holds also in the $r=4$ case if the free $k_5=0$ theory is added. The K\"ahler moduli space of $X$ is described as the complex structure moduli space of the mirror manifold $X^*$, i.e. by the odd cohomology of $X^*$. We represent the action of the symmetry generator $g$ on the toric part of the even cohomology $\tilde{H}^{\mathrm{even}}(X,\mZ) = \tilde{H}^3(X^*,\mZ)$ by a $p \times p$ matrix $A^{(G)}$, $p=2\tilde{h}^{1,1}+2$, which is determined as follows. At the Gepner point we have local coordinates $\psi_i, i=1,\dots,\tilde{h}^{1,1}(X)$. At this point there is a $\mZ_d$ monodromy 
\begin{equation}
  \label{eq:Amonodromy}
  A:(\psi_1,\dots,\psi_{\tilde{h}^{1,1}}) \to (\alpha\psi_1,\dots,\alpha^{n_{\tilde{h}^{1,1}}}\psi_{\tilde{h}^{1,1}})
\end{equation}
induced by the discrete quantum symmetry~\eqref{eq:quantum_symmetry} where $\alpha$ is a $d$th root of unity and $n_i$ are some definite integers depending on the $k_j$ with $n_1=1$. Accordingly, there is a basis of periods on the mirror manifold $X^*$
\begin{subequations}
\label{eq:Gepner_periods}
\begin{align}
  \label{eq:G.1}
  \varpi^{(G)} &= (\varpi_0,\varpi_1,\dots,\varpi_{p-1})\\
\intertext{defined by}
  \label{eq:G.2}
  \varpi_k(\psi_i) &= \varpi_0(\alpha^{k n_i}\psi_i) \qquad k=1,\dots,p-1
\end{align}
\end{subequations}
which behaves under this monodromy as
\begin{equation}
  \label{eq:Gepner monodromy}
  \varpi^{(G)} \longrightarrow A\varpi^{(G)}\textrm{ with } A^{(G)} = \left(\begin{array}{ccccc} 0&1&&&\raisebox{-2.5ex}[-2.5ex] {\Large 0}\\&\ddots&\ddots&& \\&&\ddots&\ddots&\\\raisebox{2.5ex}[-2.5ex] {\Large 0}&&&0&1\\a_{p1}&a_{p2}&\cdots&a_{p-1,p}&a_{pp} \end{array}\right)
\end{equation}
satisfying $A^d=1$. Here $\varpi_0(\psi)$ is the period obtained by analytic continuation of the fundamental period $w_0$ at large volume; i.e. $w_0$ is the unique logarithm-free solution of the Picard-Fuchs equations. $A^{(G)}$ is the matrix representing the monodromy $A$ around the Gepner point in the Gepner basis. This choice of basis is indicated by the superscript $(G)$. The entries of the bottom row satisfy $a_{pi} \in \{-1,0,1\}, i=1,\dots,p$. 

At the large volume limit we choose the integral symplectic basis of periods 
\begin{equation}
  \label{eq:large_volume_periods}
  \varpi^{(L)}=(w^{(3)},w_1^{(2)},\dots,w_{h^{1,1}}^{(2)},w_0,w_1^{(1)},\dots,w_{h^{1,1}}^{(1)})
\end{equation}
introduced in~\cite{Hosono:1995qy},~\cite{Hosono:2001eb} which is naturally associated to the intersection form given by the open string Witten index. We won't need its precise form here except the fact that the analytic continuation of $w_0$ to the Gepner point is  $\varpi_0(\psi)$. 

In general, the two bases of periods are related through analytic continuation by a transformation $M$
\begin{equation}
  \label{eq:basis transformation}
  \varpi^{(L)} = M\varpi^{(G)}
\end{equation}
which allows to transport the monodromy matrix $A^{(G)}$ from the Gepner point to the large volume where it becomes $A^{(L)} = M A^{(G)} M^{-1}$. Although there exist methods to determine $M$ we do not need its precise form. In fact, we only need to know how $\varpi_0$ and $\varpi_1$ are related to the large volume periods. This can be determined in general by passing intermediately to the conifold. The vanishing period $\Pi_v$ at the conifold can be expressed both in terms of the Gepner basis and the large volume basis which leads to the relation~\cite{Candelas:1991rm}
\begin{equation}
  \label{eq:vanishing_period}
  \Pi_v = w^{(3)} = \varpi_1 - \varpi_0
\end{equation}
where $w^{(3)}$ is the period with the $(\log)^3$ type behaviour and corresponds to the D6-brane on $X$, see below. This relation was also fundamental in relating rational boundary states in the Gepner model $(3)^5$ to bundles on the quintic in~\cite{Brunner:2000jq}. 

The distinction of the $r=4$ and the $r=5$ cases by the $\mZ_2$ factor in $G_{\textrm{GP}}$ also has an important effect on the monodromies. According to the Greene-Plesser construction, the mirror of $X = \mP^4_w[d]$ may be identified with the family of Calabi-Yau threefolds of the form $\{W=0\}/G_{\textrm{GP}}$. In the $r=4$ case, the most general form form of $W$ is
\begin{equation}
  \label{eq:generalW}
  W = \sum_{j=1}^4 a_jz_j^{k_j+2} + a_5z_5^2 + a_6z_1z_2z_3z_4z_5 + a_7(z_1z_2z_3z_4)^2 + \dots 
\end{equation}
where $\dots$ denote the remaining $G_{\textrm{GP}}$-invariant monomials of weighted degree $d$ whose precise form is irrelevant for the discussion below. Due to the $\mZ_2$ factor in $G_{\textrm{GP}}$, we have $\sum_{i=1}^4 w_i = \frac{d}{2}$ and the last term in~\eqref{eq:generalW} therefore exists only for $r=4$. In this case, the remaining monomials are $z_5$-independent.

In general, the automorphism group of $\mP^4_w/G_{\textrm{GP}}$ not only consists of the scaling symmetries $z_j \to \lambda_jz_j$ but also includes nonlinear transformations\cite{Candelas:1994hw}. In the $r=4$ case, there always exists a particular nonlinear transformation,
\begin{equation}
  \label{eq:x5transf}
  z_5 \to z_5 + Cz_1z_2z_3z_4
\end{equation}
This transformation can be used to set the coefficient $a_7$ in~\eqref{eq:generalW} to zero. Applying~\eqref{eq:x5transf}, as well as the other nonlinear transformations, if present, and then using the scaling symmetries, we may reduce~\eqref{eq:generalW} to the form
\begin{equation}
  \label{eq:Wreduced}
  W = \sum_{j=1}^4 z_j^{k_j+2} + z_5^2 - d \psi_1 z_1z_2z_3z_4z_5 + \dots
\end{equation}
This reduction is not unique. In order to have a moduli space for the complex structures of $X^*$, we have to mod out by the set of all the transformations bringing~\eqref{eq:generalW} to~\eqref{eq:Wreduced}. Modding out by the scaling symmetries preserving~\eqref{eq:Wreduced} induces an action $A$ of the quantum symmetry group $\mZ_d$ on the parameter space $\{\psi_1,\dots,\psi_{\tilde{h}^{1,1}}\}$, as discussed above. Observe that invariance under the $\mZ_2$ factor in $G_{\textrm{mirr}}$ requires that the terms $\dots$ in~\eqref{eq:generalW} only contain even powers of the $z_j$, and hence the exponents $n_i$, $i > 1$ in~\eqref{eq:Amonodromy} are even. Hence, $A^{\frac{d}{2}}$ induces $\psi_1 \to -\psi_1$, $\psi_i = \psi_i$ for $i>1$. From the expression of the holomorphic 3-form $\Omega$ in terms of $W$ one then finds that $\varpi_{i+\frac{d}{2}} = -\varpi_i$, or in other words, that $\left(A^{(G)}\right)^{\frac{d}{2}} = -\bb1$. 

Next, modding out by the nonlinear transformations also induces an action on the parameter space~\cite{Candelas:1994hw}. In particular, modding out by~\eqref{eq:x5transf} leads to
\begin{equation}
  \label{eq:Imonodromy}
  I^q:(\psi_1,\psi_2,\dots,\psi_{\tilde{h}^{1,1}}) \to (-\psi_1,\psi_2,\dots,\psi_{\tilde{h}^{1,1}})
\end{equation}
Here $q\in\mN$ indicates that this transformation in general appears as a power of a more basic nonlinear transformation $I$. Combining this result with the discussion above, we conclude that $A^{\frac{d}{2}}I^q$ acts trivially on the parameter space. The corresponding transformation of the $z_j$'s is
\begin{equation}
  \label{eq:xtransf}
  (z_1,z_2,z_3,z_4,z_5) \to ((-1)^{w_1}z_1,(-1)^{w_2}z_2,(-1)^{w_3}z_3,(-1)^{w_4}z_4,-z_5 + d\psi_1z_1z_2z_3z_4)
\end{equation}
It can be checked that this acts as $-1$ on $\Omega$ of each $X^*$ in the family, and hence is an R-symmetry. The subgroup of the automorphism group containing all of the actual symmetries, but not this R-symmetry, is generated by $A^2$ and $I$. In this way, we see that the $\mZ_2$ ``phase symmetry'' is actually a nonlinear transformation, $I^q$. By comparison to the previous discussion, we see that the analogues of $A^2$ and $I^q$ in the Gepner model are the simple currents $s^2$ and $v$. 

\section{Boundary states and bound states}
\label{sec:Boundary_states}

In this section we will argue how the D0-brane appears as a bound state built from a certain set of boundary states in the Gepner model. We begin with a short summary of those results on these boundary states, first constructed in~\cite{Recknagel:1998sb}, which will be used in the remainder of the paper. Since we are only interested in the D0-brane, we restrict ourselves to the B-type boundary states. These are labeled as $\BS{L;M;S}\equiv\BS{L_1,\dots,L_r;M;S}$ where $0\leq L_j \leq \lfloor k_j/2\rfloor$, $0\leq M \leq 2K-1$ and $S=\sum_j S_j = 0,2$. Here, and in the following $K=\lcm{k_j+2} =d$. The label $M$ is determined in terms of the boundary state labels $M_j$ of the minimal models as
\begin{equation}
  \label{eq:Definition of M}
  M = \sum_{j=1}^r w_j M_j
\end{equation}
As the generator $g$ of the discrete symmetry group $\mZ_K$ acts on $\BS{L;M;S}$ by $M \to M+2$, the set of $M$'s can be identified with the set $\mZ_K$. For fixed $L$ the states with different $(M,S)$ form an orbit under this $\mZ_{K}$ quantum symmetry. Since the two values of $S$ correspond to a brane and its anti-brane, we restrict ourselves to the states with $S=0$. We denote the set of states obtained from a given state $\BS{L_1,\dots,L_r;0;0}$ by applying $g$ to it as its $L$-orbit 
\begin{equation}
  \label{eq:L-orbit}
  \BS{L_1,\dots,L_r} \equiv \left.\left\{ g^M\BS{L_1,\dots,L_r;0;0} \right| M= 0,\dots,K-1\right\}
\end{equation}
The central object in the comparison between the Gepner and the large volume spectra is the intersection form, in particular its conformal field theory analog, the open string Witten index $I^B = {}_B\langle\!\langle \tilde{L};\widetilde{M};\widetilde{S}|(-1)^F\BS{L;M;S}$~\cite{Douglas:1999hq}. It has been determined in~\cite{Brunner:2000jq},~\cite{Kaste:2000id} for the B-type boundary states with $r$ odd to be 
\begin{subequations}
  \label{eq:B Witten index}
  \begin{align}
  \label{eq:B Witten index odd}
  I^B_{LMS,\tilde{L}\widetilde{M}\widetilde{S}} &= C(-1)^{\frac{S-\widetilde{S}}{2}}\prod_{j=1}^r \sum_{l_j=1}^{2k_j+4} \delta^{(K)}_{\frac{l}{2}+\frac{M-\widetilde{M}}{2}} n_{L_j,\tilde{L}_j}^{l_j}\\
\intertext{while for $r$ even it is}
  \label{eq:B Witten index even}
  I^B_{LMS,\tilde{L}\widetilde{M}\widetilde{S}} &= C'(-1)^{\frac{S-\widetilde{S}}{2}}\prod_{j=1}^r \sum_{l_j=1}^{2k_j+4} (-1)^{\frac{M-\widetilde{M}+l}{K}}\delta^{(\frac{K}{2})}_{\frac{l}{2}+\frac{M-\widetilde{M}}{2}} n_{L_j,\tilde{L}_j}^{l_j}
  \end{align}
\end{subequations}
where $l=\sum_j w_j l_j$ and $\delta_x^{(n)} =  1$ if $x=0 \mod n$ and zero otherwise. The $n_{L_j,\tilde{L}_j}^{l_j}$ are the annulus coefficients of the minimal models which in the case of A-type modular invariants can be identified with the $SU(2)_{k_j}$ fusion coefficients~\cite{Naka:2000he}. Due to the $\mZ_K$ symmetry the intersection form can be written as a polynomial in the generator $g$ as follows. There is a linear transformation $t_{L_j}$ which generates the different factors in~\eqref{eq:B Witten index} for $L_j\not=0$ from $n_{0,0}=(1-g^{-w_j})$ with which the intersection form~\eqref{eq:B Witten index} can be rewritten in a useful way as~\cite{Diaconescu:2000vp}
\begin{equation}
  \label{eq:t transformation}
  t_{L_j} = t_{L_j}^T = \sum_{l=-\frac{L_j}{2}}^{\frac{L_j}{2}} g_j^l
\end{equation}
and therefore
\begin{equation}
  \label{eq:simple fusion rules}
  n_{L_j,\widetilde{L}_j} = t_{L_j} n_{0,0} t_{\widetilde{L}_j}
\end{equation}
Hence starting from the boundary state $\BS{0;M;0}=\BS{0,0,0,0,0;M;0}$ we can obtain all the other boundary states by
\begin{equation}
  \label{eq:boundary states with L > 0}
  \BS{L;M;0} = \prod_{j=1}^r t_{L_j}\BS{0;M;0}
\end{equation}
The intersection form for the B-type boundary states then becomes (up to a factor of $C(-1)^{\frac{S-\widetilde{S}}{2}}$)
\begin{equation}
  \label{eq:intersection form B-type}
  I^B_{L,\widetilde{L}}(g) = \prod_{j=1}^r n_{L_j,\widetilde{L}_j}
\end{equation}
Note that in particular for the $\sum L_j=0$ states we have
\begin{equation}
  \label{eq:intersection number for L=0}
  I^B_{00}(g) = \prod_{j=1}^r (1-g^{-w_j})
\end{equation}
The number of conformal field theory moduli of such a boundary state is given~\cite{Brunner:2000jq} by the constant term in
\begin{equation}
  \label{eq:CFT dimension}
  m^{(\textrm{CFT})} = \frac{1}{2}P^B(g) - \nu
\end{equation}
where $\nu$ is determined by~\cite{Kaste:2000id}
\begin{equation}
  \label{eq:number of vacua}
  \nu=2^{\widetilde{l}}, \qquad \widetilde{l} = \begin{cases}l    & n+r \textrm{ odd}\\
                   {l-1}& n+r \textrm{ even, }l>0\\
                   0      & n+r \textrm{ even, }l=0
  \end{cases}
\end{equation}
and $l$ is the number of labels $L_j$ in $\BS{L_1,\dots,L_r}$ that satisfy $L_j = \frac{k_j}{2}$. In our cases $n=3$. Here, $P^B(g) = \prod_{j=1}^r |n_{L_j\widetilde{L}_j}|$ is a polynomial in $g$ obtained by replacing all minus signs by plus signs in each $n_{L_j,\widetilde{L}_j}$. 

Finally, we need to know the low-energy charges of these boundary states describing them as D-branes. We introduce two bases of charges $n^{(L)}$ and $n^{(G)}$ corresponding to the periods $\varpi^{(L)}$ in~\eqref{eq:large_volume_periods} and $\varpi^{(G)}$ in~\eqref{eq:Gepner_periods}, respectively, such that the BPS central charge can be expressed as
\begin{equation}
  \label{eq:BPS_central_charge}
  Z = \sum_{i=0}^{p-1} n^{(L)}_i \varpi_i^{(L)} = \sum_{i=0}^{p-1} n^{(G)}_i \varpi_i^{(G)}
\end{equation}
As mentioned at the end of Section~\ref{sec:Symmetries}, it was argued in~\cite{Brunner:2000jq} that
\begin{equation}
  \label{eq:n(D6)}
  n^{(L)}(\BS{0;0;0}) = n^{(L)}(\mathrm{D6}) = (1,0,\dots,0)
\end{equation}
The charges of the other boundary states $\BS{L;0;0}$ are then obtained from~\eqref{eq:boundary states with L > 0} by replacing $g$ by $A^{(L)}$. Similarly, the charges of $\BS{L;2M;0}$ in~\eqref{eq:L-orbit} are determined by acting with $(A^{(L)})^{-M}$ on the right of $n(\BS{L;0;0})$. In the following we will denote the representation of the product of the $t_{L_j}$ in~\eqref{eq:boundary states with L > 0} on $H^{\mathrm{even}}(X,\mZ)$ by $T^{(G)}$ and $T^{(L)}$ in the Gepner and the large volume basis, respectively. Note also that these monodromy matrices act with their inverse on the charge vectors $n$. However, due to the $\mZ_K$ symmetry we can work without taking the inverse as well, if we simultaneously shift the $M$-label of the boundary state by $M \to 2K-M$. This is, for simplicity, what we will implicitly do in the remainder of the text.

After this summary of results on boundary states in the Gepner model, we briefly review their relation to fractional brane states in the corresponding LG orbifold theory. In~\cite{Diaconescu:2000ec} it was conjectured that the restrictions of fractional brane states of the orbifold theory $\mC^5/\mZ_K$ to a Calabi-Yau hypersurface $X$ in a weighted projective space, represent the rational B-type boundary states of the Gepner model that describes the small volume limit of $X$. More precisely, the fractional brane states with the fields in the {\bf 10} of $SU(5)$ set to zero generically correspond to the Gepner model boundary states with $\sum_j L_j =0$ in~\eqref{eq:boundary states with L > 0} (see also~\cite{Govindarajan:2001vi},~\cite{Tomasiello:2001ym},~\cite{Mayr:2001as}). We will use this correspondence to present our arguments. 

The main observation is that picking a location for the D0-brane such that the $\mZ_2$ part corresponding to the trivial subtheory of the symmetry group $G_{\textrm{mirr}}$ gets broken will leave the actual phase symmetry group $\prod_{j=1}^4 p_j$ of the Gepner model unbroken. E.g. in the coordinate patch $z_1=1$ we can describe a point in a manner invariant under the scaling symmetries $z_j \to \lambda_jz_j$ by the intersection of the three hyperplanes $z_2=0$, $z_3=0$ and $z_4=0$. If $r=4$, then $W$ reduces to $z_5^2=0$. This would give us two D0-branes sitting at $z_5=0$. Since a D0-brane respects B-type boundary conditions, it is constructed as an A-type boundary state on the mirror, and should therefore respect the phase symmetries $p_j$. Roughly speaking, since $p_5$ is not a true phase symmetry, we may break it. Hence we expect to find the D0-brane in Gepner models with $r=4$.

According to~\cite{Douglas:1997de} a D0-brane sitting at the orbifold point is built from the sum over all fractional brane states in the orbifold theory. Usually, performing the blow-up would turn this D0-brane into a ``D8-brane'' wrapping the exceptional divisor. From the point of view above, however, the D0-brane is still fractional due to the residual $\mZ_2$ symmetry. Taking this into account, we have two D0-branes sitting at $z_5^2=0$ which will not lead to wrapped D-branes, but will get additional moduli which allow them to move on the restriction of the exceptional divisor to the Calabi-Yau hypersurface. Having in mind both pictures, the B-type boundary states with $\sum_j L_j =0$ and the fractional brane states at the orbifold, we proceed to show this behaviour explicitly. 

We begin with a simple observation. We look at the effect on a B-type boundary state in the $r=4$ Gepner model if an additional free theory with $k_5=0$ is added. Since the corresponding weight is $w_5=\frac{K}{2}$ this changes the $M$ and $S$ labels by $M \to M + \frac{K}{2}M_5$ and $S \to S + S_5$. Taking into account the condition that $M_5+S_5 \in 2\mZ$ we see from~\eqref{eq:B Witten index even} that $(M,S) \to (M+K,S)$ is equivalent to $(M,S) \to (M,S+2)$. This means that the set of boundary states in a given $L$-orbit of a Gepner model including a free factor consists of twice the set of boundary states in the corresponding $L$-orbit of the corresponding Gepner model without the free theory, once as branes and once as their anti-branes. This is another reflection of the two interpretations of the effect of the simple current $v$ in~\eqref{eq:GPgroup}. Viewed as a vector current, it corresponds to shifting $S \to S+2$, which is then formally equivalent to viewing it as a phase symmetry $p_5$ associated to the $k_5=0$ factor which shifts $M \to M + K$. 

Another way to see this is to look at the low-energy charges corresponding to these boundary states. At the end of Section~\ref{sec:Symmetries} we have argued that the actions of $A^{\frac{K}{2}}$ and $I^q$ are inverse to each other, corresponding to the simple current $v$, or for that matter, $p_5$. From~\eqref{eq:Gepner monodromy} and~\eqref{eq:BPS_central_charge} we see that the fact that $A^{\frac{K}{2}} = -\bb1$ has the following consequence. If we start with any boundary state in a given $L$-orbit and construct the other states by acting with $g$ as explained above, then we find after $K/2$ actions that $n(\BS{L;M+K;0}) = -n(\BS{L;M;0})$. By the above, this is the same as $n(\BS{L;M;2})$. So, the boundary states in an $L$-orbit always consist of pairs of a brane and its anti-brane. The set of boundary states $B=\{\BS{0;2l;0} | l = 0,\dots,K-1\}$ therefore corresponds to two copies of the fractional brane states in the $\mC^5/\mZ_K$ orbifold theory. Hence, in order to obtain the D0-brane we only need to sum over one half of all boundary states in $B$, i.e.
\begin{equation}
  \label{eq:2D0}
  \mathrm{2D0} = \sum_{l=0}^{\frac{K}{2}-1} \BS{0;2l;0} = \sum_{l=0}^{\frac{K}{2}-1} g^l \BS{0;0;0} = \BS{\tfrac{K}{2}-1,0,0,0,0;0;0}
\end{equation}
where the last equality follows from~\eqref{eq:boundary states with L > 0} and~\eqref{eq:t transformation}. Note that using the $\mZ_K$ symmetry we have shifted the summation by $\frac{K-1}{2}$. The 2 in front of the D0 will become clear in a moment.

As for the low-energy charges $n^{(L)}(\mathrm{2D0})$ at large volume of the boundary state $\BS{\frac{K}{2}-1,0,0,0,0;0;0}$ we find
\begin{equation}
  \label{eq:n(D0)}
  n^{(L)}(\mathrm{2D0}) = n^{(L)}(\mathrm{D6}) T_2^{(L)} = n^{(G)}(\mathrm{D6})T_2^{(G)} M^{-1}
\end{equation}
where $n^{(G)}(\mathrm{D6})$ are the low-energy charges in the Gepner basis~\eqref{eq:n(D6)} and $M$ is the transformation matrix from the Gepner basis to the large volume basis~\eqref{eq:basis transformation}. The D6-brane with charge $n^{(L)}(\mathrm{D6})=(1,0,\dots,0)$ is mirror to the D3-brane vanishing at the conifold point on the mirror, and its central charge, i.e. its period is given in terms of the periods in the Gepner basis by~\eqref{eq:vanishing_period}, whence $n^{(G)}(\mathrm{D6}) = (-1,1,0,\dots,0)$. Therefore we need to know only the two top rows of $T_2^{(G)}$. Actually we can compute the full $T_2^{(G)}$. In terms of $A^{(G)}$, using the fact that in our case $\left(A^{(G)}\right)^{\frac{K}{2}} = -\bb1$ the transformation matrix $T_2^{(G)}$ will take the following form 
\begin{equation}
  \label{eq:T_L(2D0)}
  T_2^{(G)} = \sum_{l=0}^{\frac{K}{2}-1} \left(A^{(G)}\right)^l =  \frac{2}{1-A^{(G)}}
\end{equation}
The inverse of $1-A^{(G)}$ can be computed explicitly and~\eqref{eq:T_L(2D0)} becomes
\begin{equation}
  \label{eq:T_L(2D0)2}
  T_2^{(G)} = \frac{2}{1-\sum_{i=1}^p a_{pi}}\left(\begin{array}{ccccc} 
  1-\sum_{i=2}^pa_{pi} & 1-\sum_{i=3}^pa_{pi} & \cdots & 1-a_{pp} & \hphantom{-a}1\hphantom{-a} \\
  a_{p1} & 1-\sum_{i=3}^pa_{pi} & \cdots & 1-a_{pp} & 1 \\
  a_{p1} & \sum_{i=1}^2 a_{pi} & \cdots & 1-a_{pp} & 1 \\
  \vdots & \vdots & \ddots & \vdots & \vdots \\
  a_{p1} & \sum_{i=1}^2 a_{pi} & \cdots & 1-a_{pp} & 1 \\
  a_{p1} & \sum_{i=1}^2 a_{pi} & \cdots & \sum_{i=1}^{p-1}a_{pi} & 1
  \end{array}\right)
\end{equation}
This gives us $n^{(G)}(\mathrm{D6})T_2^{(G)} = (-2,0,\dots,0)$, i.e. the charge is given by two times the fundamental period. Note that since all states appear as pairs of brane and its anti-brane we don't care about the overall sign of $n^{(L)}$ and denote the state as a brane irrespective of this sign. Now, the basis transformation matrix $M$ has the property that it leaves the fundamental period invariant, and, in our notation, changes just its position to the $(\tilde{h}^{1,1}+2)$nd entry, see~\eqref{eq:large_volume_periods}. Hence in models with $r=4$ factors the $L$-orbit $\BS{\frac{K}{2}-1,0,0,0,0}$ always contains a bound state of 2 D0-branes. 

We can get more information by looking at the number of moduli of these boundary states. From the previous discussion we see that the weights satisfy $w_1+w_2+w_3+w_4=\frac{K}{2}$ and $w_5=\frac{K}{2}$. According to the prescription given after~\eqref{eq:number of vacua} we obtain for $P^B(g)$, assuming for simplicity that $w_1=1$
\begin{eqnarray}
  \label{eq:P^B(2D0)}
  \nonumber P^B(g) &=&\left(g^{-\frac{K}{2}} + g^{-\frac{K}{2}+1} + \dots + g^{-1} + 1 + g + \dots + g^{\frac{K}{2}-1}\right)\\
         &&\left(1+g^{-w_2}\right)\left(1+g^{-w_3}\right)\left(1+g^{-w_4}\right)\left(1+g^{-\frac{K}{2}}\right)
\end{eqnarray}
The counting of the terms contributing to the constant term goes as follows: There are four terms $1\cdot1\cdot1\cdot1\cdot1$, $g^{-1}g^{-w_2}g^{-w_3}g^{-w_4}g^{-\frac{K}{2}}$, $g^{-\frac{K}{2}}\cdot1\cdot1\cdot 1\cdot g^{-\frac{K}{2}}$ and $g^{\frac{K}{2}-1}g^{-w_2}g^{-w_3}\cdot g^{-w_4}\cdot1$. Furthermore, since $1 \leq w_2,w_3,w_4 \leq \frac{K}{2}-3$ there are terms of the form $g^{w_j}\prod_{i=2}^4 g^{-\delta_{ij}w_i} \cdot 1$ and $g^{-w_j-1}\prod_{i=2}^4 g^{-(1-\delta_{ij})w_i} \cdot g^{-\frac{K}{2}}$ for $2 \leq j \leq 4$, as well as terms of the form $g^{w_j+w_k}\prod_{i=2}^4 g^{-(1-\delta_{ij})(1-\delta_{ik})w_i} \cdot 1$ and $g^{-w_j-w_k-1}\prod_{i=2}^4 g^{-(1-\delta_{ij})(1-\delta_{ik})w_i} \cdot g^{-\frac{K}{2}}$ for $2 \leq j < k \leq 4$. This makes a total of sixteen terms. Noting that $\frac{k_1}{2} = \frac{K}{2}-1$ and $\frac{k_5}{2} = 0$ we have from~\eqref{eq:CFT dimension} and~\eqref{eq:number of vacua}
\begin{align}
  \label{eq:mCFT(2D0)}
  m^{\mathrm{(CFT)}}(\mathrm{2D0}) &= 6 & \nu = 2
\end{align}
Assuming that a single D0-brane should have three moduli, this boundary state describes a bound state of two D0-branes each of which has three moduli~\cite{Fuchs:2001fd}. We will confirm this by explicit computation in the examples below. 

\section{Elliptic fibrations}
\label{sec:Elliptic_fibrations}

So far we have found a boundary state describing two D0-branes. In~\cite{Diaconescu:2000vp}, however, there was a boundary state containing only a single D0-brane. In this section we will show that this is due to the fact that the model considered there admitted an elliptic fibration. It turns out that this, together with the fact that the corresponding Gepner model has only four factors, is actually the reason for the existence of this single D0-brane. In order to give an argument for this claim we first need to review some known facts about elliptic fibrations in toric Calabi-Yau hypersurfaces. 

Elliptic fibrations generically appear when there are singular points on $X$ inherited from the ambient variety. Recall that this ambient space is a single weighted projective space $\mP^4_w$ with weights $w=(w_1,\dots,w_5)$. Suppose the weights $w_4$ and $w_5$ have a common factor $N$ and furthermore that the remaining weights add up to $N$
\begin{align}
  \label{eq:weights}
  w_1+w_2+w_3 &= N & w_4 &= aN & w_5 &= bN
\end{align}
Then the singularities are locally of the form $\mC^3/\mZ_N$, $N\geq 3$~\cite{Hosono:1995qy}. Resolving them leads to an exceptional divisor $F$ which is either a $\mP^2$ or a Hirzebruch surface $\mF_n$ for $0 \leq n \leq 12$. This divisor is equivalent to a section of an elliptic fibration. As an example consider $X=\mP^4_{1,1,1,6,9}[18]$ with $N=3$, $a=2$, $b=3$, and $F=\mP^2$ which was studied in detail in~\cite{Candelas:1994hw}. 

In an ambient variety given by a single weighted projective space there can be three different types~\cite{Klemm:1995tj}, \cite{Klemm:1998ts} of generic fibers: $\mP^2_{1,a,b}[c]$ with $(a,b) = (1,1),\,(1,2)$ or $(2,3)$ and $c=1+a+b$. Assuming furthermore that the fibration admits a Weierstrass form then it has $k$ cohomologically inequivalent sections~\cite{Berglund:1999va}, \cite{Andreas:1999ty} where $k$ depends on $(a,b)$ as follows: $k=1$ for $(a,b)=(2,3)$, $k=2$ for $(a,b)=(1,2)$ and $k=3$ for $(a,b)=(1,1)$. In the toric description, there is no distinction between the different sections. This means that the toric divisor $F$ always consists of $k$ components, i.e. it is reducible unless $(a,b)=(2,3)$. This is also seen in the fact that the number of linearly independent toric divisors $\tilde{h}^{1,1}$ is related to the number of all linearly independent divisors by $\tilde{h}^{1,1} = h^{1,1}+1 - k$. 

It has been suggested in~\cite{Berglund:1995gd} that the missing divisors can be introduced by deforming the toric polyhedron underlying the weighted projective space, namely by cutting off a corner, and thereby changing the number of complex structure deformations of the mirror. However, performing this operation amounts to destroying the LG orbifold phase in the language of Witten's phase diagram~\cite{Witten:1993yc} and therefore the comparison between the Gepner point and the large volume limit cannot be carried out any more in the known way. The reason for losing the LG orbifold phase is, that it is always associated to the simplicial triangulation of the toric polyhedron, and by cutting off a corner, the polyhedron is not simplicial anymore. 

The description of an elliptic fibration in terms of a Gepner model is as follows. First, in the symmetry group $\mZ_K$ at the Gepner point there is always a $\mZ_N$ subgroup coming from the two levels which define the singularity which acts only on the fields of the remaining three factor theories. Recall that the action of the $\mZ_K$ quantum symmetry on the corresponding LG fields is given by~\eqref{eq:quantum_symmetry}. In terms of this LG orbifold, there is a $\mC^3/\mZ_N$ orbifold inside the $\mC^5/\mZ_K$ orbifold which describes the geometry normal to the singularity. Indeed, by~\eqref{eq:weights} and defining $h=g^c$ we see that~\eqref{eq:quantum_symmetry} becomes
\begin{align}
  \label{eq:Z_N_orbifold} 
  h: \Phi_j &\longrightarrow \e{2 \pi i\frac{w_j}{N}} \Phi_j & j=1,\dots,3
\end{align}
which coincides with the $\mZ_N$ action on $\mC^3$ defined in~\cite{Diaconescu:2000dt} while the remaining fields remain unchanged.

As mentioned above, for $k=2$ ($k=3$) there are two (three) simultaneous blow-ups because there are also non-toric divisors. Since they cannot be treated by an orbifold subtheory we will not find the D0-brane in these elliptic fibrations. In fact, one can even go further in these cases. First, the $k=3$ case does not appear in an $r=4$ Gepner model. Second, in the $k=2$ case, the boundary states corresponding to bundles on the base $F$ of the fibration do not appear among the $\sum_j L_j =0$ states, because the latter have $\nu=1$ while the former have $\nu=2$, accounting for the reducibility of the divisor $F$~\cite{Scheidegger:2001ab}. 

Since there is this $\mC^3/\mZ_N$ orbifold subtheory, the fractional branes of this subtheory will be included in the set of fractional branes on the full $\mC^5/\mZ_K$ orbifold theory. In particular, the sum of the $N$ fractional branes of the subtheory will be a D0-brane sitting at the singularity $\mC^3/\mZ_N$. Now, in the Gepner model, these fractional branes correspond to part of the $\sum_j L_j = 0$ boundary states which are related to each other by the action of $g$. This can also be seen from the large volume point of view. The set of sheaves on the resolution of the $\mC^3/\mZ_N$ orbifold are contained in the set of the sheaves coming from the $\sum_j L_j = 0$ boundary states. E.g. the sheaves on the exceptional $\mP^2$ sitting inside $\sO_{\mP^2}(-3)$ in~\cite{Diaconescu:2000dt},~\cite{Douglas:2000qw} form part of the $\sum_j L_j = 0$ sheaves in the Gepner model of the family $\mP^4_{1,1,1,6,9}[18]$ in~\cite{Diaconescu:2000vp}. Other examples of D-branes on orbifolds of the form $\mC^3/\mZ_N$ are discussed in~\cite{Mukhopadhyay:2001sr} and~\cite{DelaOssa:2001xk}. It can be shown in general~\cite{Scheidegger:2001ab} that these D-branes appear in the spectrum of the corresponding elliptic fibration. 

Performing the sum over the fractional brane states amounts to forming the bound state 
\begin{equation}
  \label{eq:D0}
  \mathrm{D0} = \sum_{l=0}^{N-1} g^l \BS{0;2N;0} = \BS{N-1,0,0,0,0;2N;0}
\end{equation}
therefore we expect to find a D0-brane brane in the $L$-orbit $\BS{N-1,0,0,0,0}$. That $n^{(L)}(\BS{N-1,0,0,0,0;2N;0}) = n^{(L)}(\mathrm{D0})$ is shown as follows. From~\eqref{eq:D0} we read off that
\begin{equation}
  \label{eq:TG(D0)}
  T_1^{(G)} = \sum_{l=0}^{N-1} \left(A^{(G)}\right)^{l+N} = \frac{1-\left(A^{(G)}\right)^N}{1-A^{(G)}}\left(A^{(G)}\right)^N
\end{equation}
and therefore the charge of $n^{(L)}(\BS{N-1,0,0,0,0;2N;0})$ is 
\begin{equation}
  \label{eq:n(K-1)}
  n^{(G)}(\mathrm{D6}) \frac{1}{1-A^{(G)}}\left(A^{(G)}\right)^N\left(1-\left(A^{(G)}\right)^N\right)M^{-1}
\end{equation}
Since we have $d=cN$ and also $d=K$ we obtain the following relations
\begin{eqnarray}
  \label{eq:relationsN+}
  \sum_{l=0}^{c-1} g^{lN} = \frac{1-g^{cN}}{1-g} & = & 0\\
  \label{eq:relationsN-}
  \sum_{l=0}^{c-1} (-1)^l g^{lN} = \frac{1-(-g)^{cN}}{1-g} & = & 0\\
  \label{eq:relationsK}
  1 + g^{\frac{K}{2}} &=& 0
\end{eqnarray}
It follows that $g^N(1-g^N) = 1$ hence 
\begin{equation}
  T_1^{(G)} = \frac{1}{1-A^{(G)}} = \frac{1}{2}T_2^{(G)}
\end{equation}
which by comparing to~\eqref{eq:T_L(2D0)} explains why~\eqref{eq:TG(D0)} gives half the charge of the 2D0-brane state, i.e. we have a single D0-brane. We will confirm this in the examples below. 

Next, we determine the number of moduli for the $L$-orbit $\BS{N-1,0,0,0,0}$. The weights of an elliptic fibration with $(a,b) = (2,3)$ satisfy $w_1+w_2+w_3=N$, $w_4=2N$ and $w_5=3N$. According to the prescription given after~\eqref{eq:number of vacua} we obtain for $P^B(g)$, assuming for simplicity that $w_1=1$
\begin{eqnarray}
  \label{eq:P^B}
  \nonumber P^B(g) &=&\left(g^{-N} + g^{-N+1} + \dots + g^{-1} + 1 + g + \dots + g^{N-1}\right)\\
         &&\left(1+g^{-w_2}\right)\left(1+g^{-w_3}\right)\left(1+g^{-2N}\right)\left(1+g^{-3N}\right)
\end{eqnarray}
From this one can see that the following terms contribute to the constant term: $1\cdot1\cdot1\cdot1\cdot1$, $g^{-w_1}g^{-w_2}g^{-w_3}g^{-2N}g^{-3N}$, $g^{-N}\cdot1\cdot1\cdot g^{-2N}g^{-3N}$ and $g^{N-1}g^{-w_2}g^{-w_3}\cdot1\cdot1$. Furthermore, since $1 \leq w_2,w_3 \leq N-2$ there are terms of the form $g^{-w_3-1}g^{-w_2}\cdot1\cdot g^{-2N}g^{-3N}$, $g^{-w_2-1}\cdot1\cdot g^{-w_3} g^{-2N}g^{-3N}$, $g^{w_2}g^{-w_2}\cdot 1\cdot 1 \cdot 1$ and $g^{w_3}\cdot 1\cdot g^{-w_3} \cdot 1 \cdot 1$. Therefore, there are always eight terms in total and~\eqref{eq:CFT dimension} yields then
\begin{equation}
  \label{eq:mCFT(D0)}
  m^{\mathrm{(CFT)}}(\mathrm{D0}) = 3
\end{equation}
moduli in complete agreement with the fact that one expects this number from geometry.

Independently, we can construct all the boundary states in a Gepner model by explicitly computing the matrix $A^{(L)}$ using the method of~\cite{Mayr:2001as}. Given this list we can look for those which have the charges corresponding to D0-branes. This has been compiled in table~\ref{tab:D0-branes}. We see that the double D0-brane boundary state appears precisely for those models that have $r=4$. The models which allow for an elliptic fibration with a single section have a boundary state corresponding to a single D0-brane. 
\TABLE{
\def\Dph{\hphantom{\mathrm{2D0+2D6}}}
\begin{footnotesize}
$
  \begin{array}{|c|c|c|c|c|c|c|c|c|}
    \hline
      &   &  & \mathrm{2D0} & \mathrm{D0} & \mathrm{2D0+2D6} & \mathrm{4D0+2D6} & \mathrm{D0+D6} & \mathrm{2D0+D6}\\
    m^{(\mathrm{CFT})} & & & 6 & 3 & 6 & 14 & 3 & 6\\
    \nu &\hphantom{5555} &\hphantom{5555} & 2 & 1 & 2 & 2 & 1 & 1 \\
    X & r & k &\Dph &\Dph & & &\Dph &\Dph \\
    \hline
    \mP^4_{1,1,1,1,1}[5]&5&\textrm{--}& \textrm{--} & \textrm{--} & \textrm{--} & \textrm{--} & \textrm{--} & \textrm{--} \\
    \hline
    \mP^4_{1,1,1,1,2}[6]&5&\textrm{--} & \textrm{--} & \textrm{--} & \textrm{--} & \textrm{--} & \textrm{--} & \textrm{--} \\
    \hline
    \mP^4_{1,1,1,1,4}[8]&4&\textrm{--}&\BS{3,0}&\textrm{--}&\BS{3,0}&\BS{3,1}&\textrm{--}&\BS{2,0}\\
    \hline
    \mP^4_{1,1,1,2,5}[10]&4&\textrm{--}&\BS{4,0}&\BS{0,1}^*&\BS{4,0}&\BS{4,1}&\BS{0,1}^*&\BS{3,0}\\
    \hline
    \mP^4_{1,1,2,2,2}[8]&5&\textrm{--}& \textrm{--} & \textrm{--} & \textrm{--} & \textrm{--} & \textrm{--} & \textrm{--} \\
    \hline
    \mP^4_{1,1,2,2,6}[12]&4&\textrm{--}&\BS{5,0}&\textrm{--}&\BS{5,0}&\BS{5,1}&\textrm{--}&\BS{4,0}\\
    \hline
    \mP^4_{1,2,2,3,4}[12]&5&\textrm{--}& \textrm{--} & \textrm{--} & \textrm{--} & \textrm{--} & \textrm{--} & \textrm{--} \\
    \hline
    \mP^4_{1,2,2,2,7}[14]&4&\textrm{--}&\BS{6,0}&\BS{0,2}&\BS{6,0}&\textrm{--}&\BS{0,2}&\BS{5,0}\\
    \hline
    \mP^4_{1,1,1,6,9}[18]&4&1&\BS{8,0}&\BS{2,0}&\BS{8,0}&\BS{8,1}&\BS{2,0}&\BS{7,0}\\
    \hline
    \mP^4_{1,1,1,3,6}[12]&4&2&\BS{5,0}&\textrm{--}&\BS{5,0}&\BS{5,1}&\textrm{--}&\BS{4,0}\\
    \hline
    \mP^4_{1,2,3,3,3}[12]&5&\textrm{--}& \textrm{--} & \textrm{--} & \textrm{--} & \textrm{--} & \textrm{--} & \textrm{--} \\
    \hline
    \mP^4_{1,3,3,3,5}[15]&5&\textrm{--}& \textrm{--} & \textrm{--} & \textrm{--} & \textrm{--} & \textrm{--} & \textrm{--} \\
    \hline
    \mP^4_{1,2,3,3,9}[18]&4&\textrm{--}&\BS{8,0}&\BS{0,3}&\BS{8,0}&\textrm{--}&\BS{0,3}&\BS{7,0}\\
    \hline
    \mP^4_{1,1,2,8,12}[24]&4&1&\BS{11,0}&\BS{3,0}&\BS{11,0}&\BS{11,1}&\BS{3,0}&\BS{10,0}\\
    \hline
    \mP^4_{1,1,1,3,3}[9]&5&3& \textrm{--} & \textrm{--} & \textrm{--} & \textrm{--} & \textrm{--} & \textrm{--} \\
    \hline
    \mP^4_{1,1,2,4,8}[16]&4&2&\BS{7,0}&\textrm{--}&\BS{7,0}&\BS{7,1}&\textrm{--}&\BS{6,0}\\
    \hline
    \mP^4_{1,2,2,10,15}[30]&4&1&\BS{14,0}&\BS{4,0}&\BS{14,0}&\textrm{--}&\BS{4,0}&\BS{13,0}\\
    \hline
    \mP^4_{1,1,2,4,4}[12]&5&3& \textrm{--} & \textrm{--} & \textrm{--} & \textrm{--} & \textrm{--} & \textrm{--} \\
    \hline
  \end{array}
$
\end{footnotesize}
\caption{\label{tab:D0-branes} The boundary states corresponding to D0-branes for all Fermat hypersurfaces $X$ with $h^{1,1} \leq 4$. Only the $L_1$ and $L_2$ labels are indicated, the others being zero except for those marked with a * where $L_1$ and $L_4$ are displayed. The meaning of the last four columns will become clear in Section~\ref{sec:Lines}.}
}
Inspecting the table we find that there are single D0-brane states in models which do not admit an elliptic fibration. In these cases it turns out to be more difficult to display the symmetry group leading to an orbifold subtheory and to relate it to the geometry. In some Gepner models there can be another divisor $D$ which has the topology of a rational surface but does not come from a blow-up of a singularity and hence is not exceptional. Nevertheless, there are D4-branes wrapped on it which carry the same charges as in the exceptional case~\cite{Scheidegger:2001ab}. Taking bound states of them in the same way as we have done above then yields a D0-brane in the spectrum. This is the case for $\mP^4_{1,2,2,2,7}[14]$ and $\mP^4_{1,2,3,3,9}[18]$ in table~\ref{tab:D0-branes}. For the D0-brane in $\mP^4_{1,1,1,2,5}[10]$ which was also noticed in~\cite{Scheidegger:2000ed} we do not have a geometric interpretation.

\section{Walls of marginal stability}
\label{sec:Lines}

In this section we consider bound states of D0- and D6-branes. In the large volume limit, when the Calabi-Yau manifold can be approximated by a flat space, unbroken supersymmetry requires that a D$p$-D$q$-brane bound state can only exist if $p = q \mod 4$~\cite{Polchinski:1996na}. Hence, a bound state of a D0- and a D6-brane is unstable and cannot exist. This can actually be shown by computing the static force between them which turns out to be repulsive~\cite{Lifschytz:1996iq}. The possibility of having supersymmetric bound states of D0- and D6-branes when a very large $B$-field is turned on was recently considered in~\cite{Mihailescu:2000dn}, \cite{Witten:2000mf}. The issue of the stability of such bound states was also studied from a different point of view in~\cite{Denef:2001ix}.

It is interesting to look at the boundary states obtained from $\BS{\frac{K}{2}-1,0,0,0,0;0;0}$ and $\BS{N-1,0,0,0,0;2N;0}$ by acting with $g$ on them. We will show that the resulting boundary states $\BS{\frac{K}{2}-1,0,0,0,0;2;0}$ and $\BS{N-1,0,0,0,0;2N+2;0}$ will correspond to bound states of D0- and D6-branes. We first look at the charges
\begin{equation}
  \label{eq:n(D0D6)}
  n^{(L)}(s(\mathrm{D0+D6})) = n^{(L)}(\mathrm{D6}) T_s^{(L)} A^{(L)} = n^{(G)}(\mathrm{D6}) T_s^{(G)} A^{(G)} M^{-1}
\end{equation}
for $s=1,2$. By using the previous results, \eqref{eq:T_L(2D0)}, \eqref{eq:T_L(2D0)2}, \eqref{eq:TG(D0)} and~\eqref{eq:Gepner monodromy}, we find
\begin{equation}
  \label{eq:n(D0D6)2}
  n^{(L)}(s(\mathrm{D0+D6})) = (0,-s,0,\dots,0) M^{-1}
\end{equation}
Therefore we need to know the second row of $M^{-1}$, or in other words how $\varpi_1$ is related to the periods at large volume. We have already given the answer in~\eqref{eq:vanishing_period}. Therefore
\begin{equation}
  \label{eq:varpi_1}
  \varpi_1 = w^{(3)} + w_0
\end{equation}
and inserting this into~\eqref{eq:n(D0D6)2} yields the large volume charges of the boundary states $\BS{\frac{K}{2}-1,0,0,0,0;2;0}$ and $\BS{N-1,0,0,0,0;2N+2;0}$ to be $(-2,0,\dots,0,-2,0,\dots,0)$ and $(-1,0,\dots,0,-1,0,\dots,0)$ for $s=2$ and $s=1$, respectively. Hence they correspond to bound states of $s$ D0- and $s$ D6-branes as was implied by the name of the state we have given in~\eqref{eq:n(D0D6)}. Since these states are in the same $L$-orbit as the D0-branes, they have the same number of moduli as in~\eqref{eq:mCFT(2D0)} and~\eqref{eq:mCFT(D0)}, respectively. This agrees with the results of the direct computation in the columns 2D0+2D6 and D0+D6 of table~\ref{tab:D0-branes}.

We can use the relation~\eqref{eq:varpi_1} between the large volume periods and the Gepner periods to produce more such bound states of D0- and D6-branes by taking linear combinations of the two pairs of states that we have found, D0 and D0+D6 for $s=1$ and 2D0 and 2D0+2D6 for $s=2$. Taking linear combinations means applying a further transformation $T_\mathrm{lin}$ to the charge vector $n^{(G)}(\mathrm{D6})T_s^{(G)} = (-s,0,\dots,0)$ such that the transformed charge vector has non-zero entries in the first two components. From the form of $A^{(G)}$ in~\eqref{eq:Gepner monodromy} we can see that there is only one possibility, namely by acting with $T_{\mathrm{lin}}=(1+g)$ on this charge vector. This action can be achieved in two ways. The first one is to note that
\begin{equation}
  \label{eq:1+g}
  \frac{1+g}{1-g} = \frac{2}{1-g} -1 = \sum_{l=1}^{\frac{K}{2}-1}g^l
\end{equation}
The left-hand side of~\eqref{eq:1+g} is $\frac{1}{2}T_2^{(G)}(1+A^{(G)})$ and can be interpreted as the bound state of a D0-brane and a D0-D6-brane bound state. Note however, that neither of them appears in the spectrum. The middle part, on the other hand, allows for an interpretation as a bound state of the anti-D6-brane and the 2D0-2D6-brane bound state we have found above. The right-hand side is the sum over $\frac{K}{2}-2$ fractional brane states sitting at the $\mZ_K$ orbifold singularity. Therefore we will find a bound state with charges $(-1,-1,0,\dots,0)M^{-1}=(-1,0,\dots,0,-2,0,\dots,0)$ in the $L$-orbit $\BS{\frac{K}{2}-2,0,0,0,0}$. The second way is to note from~\eqref{eq:boundary states with L > 0} that if $w_1=w_2=1$ we can take the transformation $T_2^{(G)}$ for the first factor and $(1+A^{(G)})$ for the second factor. This will then yield a boundary state with large volume charges $(-2,0,\dots,0,-4,0,\dots,0)$ in the $L$-orbit $\BS{\frac{K}{2}-1,1,0,0,0}$. By a similar counting argument as in the previous sections we can then show that the first of these has 6 moduli, while the second has 14 moduli and $\nu=2$. We can again confirm these results by comparing with the examples given in table~\ref{tab:D0-branes} in the columns 2D0+D6 and 4D0+2D6.

Finally, we like to point out a second possibility to relate the states corresponding to the D0-brane to those corresponding to the bound state of a D0- and a D6-brane. For this purpose we generalize an argument given in~\cite{Andreas:2000sj} for a particular model to all Gepner models. By the standard argument of~\cite{Diaconescu:2000vp} the comparison at large volume of the central charges~\eqref{eq:BPS_central_charge} corresponding to a state with charges
\begin{equation}
  \label{eq:n}
  n^{(L)}=\left(n_6, n_4^{J_1},\dots,n_4^{J_{h^{1,1}}},n_0,n_2^{C_1},\dots,n_2^{C_{h^{1,1}}}\right) \in H^{\mathrm{even}}(X,\mZ) 
\end{equation}
yields the Chern characters of the associated sheaf $\sF$ 
\begin{eqnarray}
  \label{eq:rank D6}
  \rank(\sF) &=& n_6 \\
  \label{eq:ch1 D6}
  \chern_1(\sF) &=& \sum_{i=1}^{h^{1,1}} n_4^{J_i}\;J_i\\
  \label{eq:ch2 D6}
  \chern_2(\sF) &=& \sum_{i=1}^{h^{1,1}} \left(n_2^{C_i} + A_{ij} n_4^{J_j}\right)\;C_i\\
  \label{eq:ch3 D6}
  \chern_3(\sF) &=& -n_0-\frac{1}{12}\sum_{i=1}^{h^{1,1}} n_4^{J_i}\;\ch_2\cdot J_i
\end{eqnarray}
where the $J_i$ and the $C_i$ form a basis for $H^2(X,\mZ)$ and $H^4(X,\mZ)$ respectively, satisfying $C_i\cdot J_j = \delta_{ij}$ and $A_{ij} = \frac{1}{2} J_i\cdot J_i\cdot J_j \mod \mZ$~\cite{Hosono:1995ax}. According to~\cite{Horja:1999ab} the monodromy $T_{\mathrm{con}}$ about the conifold locus in the complex structure moduli space of the mirror $X^*$ corresponds to the automorphism of the derived category of bounded complexes of coherent sheaves on $X$ whose effect on the cohomology can be described by
\begin{equation}
  \label{eq:Tcon}
  T_{\mathrm{con}}: \gamma \longrightarrow \gamma - \left(\int_X\gamma\wedge\todd(X)\right) \cdot 1_X \qquad \gamma \in H^{\textrm{even}}(X,\mZ)
\end{equation}
which corresponds to a change in the topological invariants of the sheaf $\sF$ 
\begin{equation}
  \chern(\sF) \longrightarrow \chern(\sF) -\frac{\chern_1(\sF)\ch_2(X)}{12}+\chern_3(\sF)
\end{equation}
Now from~\eqref{eq:rank D6} and~\eqref{eq:ch3 D6} we have
\begin{equation}
  \frac{\chern_1(\sF)\ch_2(X)}{12}+\chern_3(\sF) = \frac{n_4^{J_i}\ch_2\cdot J_i}{12} - n_0 - \frac{n_4^{J_i}\ch_2\cdot J_i}{12} = -n_0
\end{equation}
hence we get that
\begin{equation}
  \label{eq:n6->n_6+n_0}
  n_6 \longrightarrow n_6 + n_0
\end{equation}
Now recall that the monodromy matrix about the conifold locus in the large volume basis is
\begin{equation}
  T_{\mathrm{con}}^{(L)} = \bb1 - E_{1,h^{1,1}+2}
\end{equation}
where $E_{ij}$ is the matrix with zeroes everywhere except at the $(i,j)$-th entry. Comparing the last two results, we see that the effect of $T_{\mathrm{con}}$ on the charge vector $n^{(L)}$ is given by $\left(T_{\mathrm{con}}^{(L)}\right)^{-1}$. Hence we see that the states $\BS{\frac{K}{2}-1,0,0,0,0;0;0}$ and $\BS{N-1,0,0,0,0;2N;0}$ can also be related by a monodromy coming from a loop around the conifold locus to the states $\BS{\frac{K}{2}-1,0,0,0,0;2;0}$ and $\BS{N-1,0,0,0,0;2N+2;0}$. This is not surprising since the derivation of~\eqref{eq:varpi_1} or~\eqref{eq:vanishing_period} in~\cite{Candelas:1991rm} is precisely related to such a loop.

From~\eqref{eq:n6->n_6+n_0} we see that if we encircle the conifold locus, we could produce bound states with arbitrarily large D6-brane charge. That such states do not exist in the spectrum is already clear from the finiteness of the construction of the boundary states. However, there are no states consisting of $s$ D0-branes or of $s$ D0-D6 brane bound states with $s > 2$. Furthermore, our analysis shows that there are no states with more D6-branes than D0-branes. This implies that there should be a wall of marginal stability through the conifold locus on which not only those states with large $n^{(L)}(\mathrm{D6})$ decay but also those with small values of D6-brane charge. For example, if we start with a D0-brane and go once around the conifold locus we obtain a D0-D6 brane bound state. Going once more around it, this state must decay into other states. This is very similar to the situation in the non-compact $\sO_{\mP^2}(-3)$ example studied in~\cite{Douglas:2000qw}. 

In~\cite{Aspinwall:2001zq} it was conjectured on the basis of~\eqref{eq:Tcon} that a D0-brane undergoes a monodromy if and only if it is transported around the conifold locus. Since the Landau-Ginzburg phase and the geometric phase in a Fermat hypersurface are always separated by the phase boundary containing to the conifold locus, we know that a D0-brane will be affected when transporting it from the Gepner point to the large volume and vice versa. Therefore the D0-brane at large volume seems not to be the same as the D0-brane at the Gepner point. This has been pointed out~\cite{Aspinwall:2001pu} where also a mathematical description of these bound states and the monodromy transform in terms of complexes of sheaves has been given. 

\section{Conclusions}
\label{sec:Conclusions}

Based on symmetry arguments we have given very simple criteria for the existence of states corresponding to D0-branes at the Gepner point of the K\"ahler moduli space of a Calabi-Yau manifold given as a Fermat hypersurface in a weighted projective space. If the corresponding Gepner model consists only of four minimal model subtheories then there is always a boundary state containing 2 D0-branes. If the Calabi-Yau manifold is in addition an elliptic fibration with a single section, then there is a state describing a single D0-brane. We also pointed out that there are other possibilities for finding a single D0-brane at the Gepner point. We have also argued that depending on the model there are at most four different types of bound states of D0- and D6-branes. 

It should be possible to derive these results completely in the framework of quiver gauge theories. For this purpose it will be necessary to understand the fields in the {\bf 10} of the $SU(5)$ that have been set to zero. Those are related to deformations of the corresponding sheaves $\sF$~\cite{Diaconescu:2000ec}. These are characterized by the group $\Ext^1(\End \sF)$. A few general comments on the relation to bound states have been given in~\cite{Douglas:2000ah}.

Finally, we have argued that in addition to the wall of marginal stability associated to non-supersymmetric D0-D6 brane bound state there is a wall coming from the monodromy around the conifold locus. We would like to point out that their origin has a very different physical nature. It would be interesting to know whether this is reflected in a direct and explicit description of these walls. 

\acknowledgments 
I would like to thank S. Theisen for many improvements of the manuscript and the referee for pointing out some misconceptions. I am grateful to M. Kreuzer for valuable discussions.

\end{document}